# Effect of Surface Treatment on High-Temperature Oxidation Behavior of IN 713C

*Wojciech J. Nowak and Bartek Wierzba*

The effects of surface preparations on oxidation kinetics and oxide scale morphology for the commercially available Ni-based superalloy IN 713C have been investigated. The ground and polished samples were exposed in air at 800-1100 °C. The ground specimens were found to demonstrate lower oxidation kinetics compared to those after polishing. The grinding also affected the oxide scale morphology, resulting in a protective alumina scale, while the polished samples developed Ni-/Cr-rich mixed oxides on the surface. Better oxidation resistance of the ground surfaces is related to a higher concentration of defects in the near-surface region introduced by cold working. These defects facilitate the outward transport of the scale-forming element Al and thus are beneficial for protective oxidation. The oxidation mechanism at lower temperatures was introduced. The model based on the generalized Darken method and multiphase approximation was proposed.

## 1. Introduction

Over the decades, high-temperature metallic materials such as Ni-based alloys have been optimized to provide high strength at room as well as at elevated temperatures (Ref 1). The latter is achieved by alloying with aluminum to form stable strengthening precipitates such as $\gamma'$. Moreover, the latter results in protective alumina scales formation on the alloy surface at high temperatures. However, current manufacturing methods limit the Al content in the alloys due to complications related to, e.g., casting, welding, forming, etc. Al and/or Cr are the elements forming protective oxide scales (chromia and alumina, respectively) during exposure at high temperatures in oxidizing atmospheres (Ref 2, 3). The protective oxide scale reduced the component wall thickness loss by preventing the oxidation of the base alloy components such as Fe or Ni which form rapidly growing oxides. Additionally, compact protective oxide scales hinder the penetration of carbon, nitrogen or sulfur, thereby preventing carburization or sulfidation (Ref 4, 5) of the materials exposed in mixed corrosive environments. Due to the limited Cr and/or Al content, numerous high-temperature construction materials are termed as "marginal" alumina or chromia formers (Ref 3). This implies that the ability of the alloy to form protective scales strongly depends on exposure conditions, i.e., temperature, gas composition, heat treatment, surface preparation, etc. Cr is known to be a strong carbide and sulfide former (Ref 1); therefore, small amounts of Cr in the alloys decrease their resistance against carburization and sulfidation. An alternative method to provide resistance against environmental reactions (oxidation, nitridation, carburization or sulfidation) is the application of protective coatings, such as MCrAlY (Ref 6) (where M is mainly Ni or Co) or β-NiAl (Ref 7, 8) which are mainly alumina-forming materials. However, coating production is generally costly, is time-consuming and may detrimentally affect the substrate material via interdiffusion processes resulting in microstructural degradation of the alloy microstructure, void formation, interface embrittlement, coating exfoliation, etc. (Ref 9, 10). Another way to promote protective external oxide scaling on high-temperature alloys without changing the material composition and/or manufacturing technology is mechanical surface treatment proposed in the present work.

After the formation of a continuous oxide scale, the reaction kinetics is controlled by the slowest transport process, i.e., diffusion of the reacting species through the oxide scale. To maintain the external scaling process, the transport of the scale-forming element toward the scale–alloy interface needs to be sufficiently fast. At temperatures within the application range of the high-temperature materials, diffusion in metals occurs over the grain volume or grain surface (at the grain boundaries). Volume diffusion has generally a much higher activation energy compared to that of surface diffusion (Ref 11). The volume-to-surface diffusion ratio is thus given by the equation:

$$\frac{D_V}{D_S} = \text{const} \cdot e^{\left(-\frac{E_V - E_S}{RT}\right)}, \qquad (\text{Eq 1})$$

where $D_V$ and $D_S$ are volume and surface diffusion coefficients, respectively, $E_V$ and $E_S$—activation energies of volume and surface diffusions, respectively, $R$—gas constant, and $T$—temperature.

The grain boundaries are regarded as crystallographic structure defects; therefore, in the present work it is proposed that the defects introduced into the near-surface region of the alloy play a similar role as grain boundaries in terms of element diffusion.

The effect of mechanical surface preparation on the oxidation morphology has been investigated in a number of experimental studies. Giggins and Pettit (Ref 12) investigated

**Wojciech J. Nowak** and **Bartek Wierzba**, Department of Materials Science, Faculty of Mechanical Engineering and Aeronautics, Rzeszow University of Technology, al. Powstancow Warszawy 12, 35-959 Rzeszow, Poland. Contact e-mail: w.nowak@prz.edu.pl.

the influence of grain size and surface deformation on the selective chromia scale formation on Ni-Cr alloys. The authors observed that the oxidation behavior of the grit-blasted samples was similar to that of the fine structured materials. This phenomenon was explained by the formation of fine grains via the recrystallization of the cold-worked subsurface layer during exposure at elevated temperatures. The effect of grain size and deformation on selective oxidation of chromium was observed up to 30 wt.% of Cr which was found to be associated with enhanced chromium diffusion over the grain boundaries.

The effect of grain boundary density, introduced into the material by cold working with subsequent recrystallization, on the oxidation and carburization of Alloy 800 has been investigated by Leistikow et al. (Ref 13). The authors found that samples with higher grain boundary densities exhibited the lowest mass gain. The tendency for internal oxidation increased with decreasing grain boundary density. This observation corroborates the role of grain boundaries as fast-circuit diffusion paths for the scale-forming element toward the alloy surface. From the crystallographic point of view, grain boundaries present in the polycrystalline materials are also viewed as defects.

In the recent study by Sudbrack et al. (Ref 14), the effect of surface preparation of two single-crystal Ni-based superalloys LDS-1101+HF and CMSX-4+Y on the oxidation behavior has been investigated. The surface finish of the samples was as follows: low-stress grinding, polishing up to 9 and 3 μm and electrochemical thinning. The authors established a clear correlation between the resulting oxide scale morphology and the surface preparation.

The aim of the present work is to investigate the effect of surface preparation on the oxidation resistance of Ni-based superalloy IN 713C between 800 and 1100 °C exposed in laboratory air. A model based on the generalized Darken method and multiphase approximation is to describe internal oxidation. The two-phase zone (internal oxidation) is mathematically described.

## 2. Experimental

Disk-shaped coupons 20 mm in diameter and of 2 mm thickness were machined from a cast rod Ni-based superalloy IN 713C (Ni—bal., Cr—14, Ta—2, Mo—4, Nb—5.1, Al—6, Ti—1, C—0.2, B—0.01, in wt.%). The sample surfaces were prepared using two different surface finishing techniques: polishing up to 1 μm and grinding with 80-grit SiC paper.

The surface roughness and topography of the samples were characterized using the profilometer HOMMEL as well as scanning electron microscopy (SEM).

The as-prepared samples were oxidized for 48 h in laboratory air 800, 950 and 1100 °C in the Carbolite STF 16/450 tube furnace. The temperature 800 °C was selected as a baseline as it is close to the actual operating temperature of numerous Ni-based superalloys.

In the present work, oxidation tests were carried out at 950 and 1100 °C to elucidate the effect of higher temperature. The samples were weighed before and after oxidation exposure using microbalance RADWAG WAA 100/C/1 with 0.1 mg accuracy to determine the oxidation kinetics.

Phase analyses of the oxidation products of selected samples were performed using an x-ray diffractometer Miniflex II made by Rigaku. As an x-ray source filtered copper lamp (CuK$_\alpha$, $\lambda = 0.1542$ nm) with a voltage of 40 kV was used. The $2\theta$ angle range varied between 20 and 120°, and step size was 0.02°/3 s. Phase composition was determined using the Powder Diffraction File (PDF) developed and issued by the ICDD (the International Center for Diffraction Data).

The exposed samples were characterized using glow discharge optical emission spectrometry (GD-OES). The GD-OES depth profiles were quantified using the procedure described in references (Ref 15-17). After GD-OES depth profiling, the samples were gold-sputtered to impart electrical conductivity, electrochemically plated with nickel and subsequently mounted in epoxy resin. Metallographic cross sections of the oxidized alloy specimens as well as of the as-cast material were prepared by a series of grinding and polishing steps, the final step being fine polishing with the $SiO_2$ suspension with the 0.25-μm granulation. The cross sections were analyzed using a Nikon Epiphot 300 optical microscope and a Hitachi S3400N scanning electron microscope (SEM).

## 3. Oxidation Modeling

The internal oxidation model is based on the generalized Darken method with drift determination and multiphase approximation for diffusing species. The fundamental principle of the model is the mass conservation law for the average amount of each component in the ternary diffusion couple, e.g., Ni-Cr-O.

$$\frac{\partial \bar{c}_i}{\partial t} = -\frac{\partial \bar{J}_i}{\partial x}, \quad i = 1, 2, 3. \tag{Eq 2}$$

The average concentration can be defined from the Gibbs concentration triangle, where the values of the phase ratio $f^\alpha$ and $f^\beta$ are known ($f^\alpha + f^\beta = 1$):

$$\bar{c}_i = f^\alpha c_i^\alpha + f^\beta c_i^\beta, \tag{Eq 3}$$

where $\bar{J}_i$ is the average component flux; $c_i^\alpha$ and $c_i^\beta$ are the phase fractions of the phases $\alpha$ and $\beta$, respectively.

$$\bar{J}_i = f^\alpha J_i^\alpha + f^\beta J_i^\beta, \quad i = 1, 2, 3. \tag{Eq 4}$$

The fluxes in the phases $\alpha$ and $\beta$ are defined by using Darken's method (Ref 18):

$$J_i^\alpha = -D_i^\alpha \frac{\partial c_i^\alpha}{\partial x} + c_i^\alpha v^\alpha, \quad i = 1, 2, 3 \tag{Eq 5}$$

and

$$J_i^\beta = -D_i^\beta \frac{\partial c_i^\beta}{\partial x} + c_i^\beta v^\beta, \quad i = 1, 2, 3, \tag{Eq 6}$$

where $D_i^\alpha$ and $D_i^\beta$ are the intrinsic diffusion coefficients of the components. The drift velocity can be calculated from the volume continuity equation in each individual phase:

$$v^\alpha = \frac{1}{c^\alpha} \sum_{i=1}^{3} D_i^\alpha \frac{\partial c_i^\alpha}{\partial x} \quad \text{and} \quad v^\beta = \frac{1}{c^\beta} \sum_{i=1}^{3} D_i^\beta \frac{\partial c_i^\beta}{\partial x}. \tag{Eq 7}$$

Finally, the mass conservation for the average concentration in a ternary system can be rewritten in the following form:

$$\frac{\partial \bar{c}_i}{\partial t} = -\frac{\partial}{\partial x}\left(-f^\alpha D_i^\alpha \frac{\partial c_i^\alpha}{\partial x} - f^\beta D_i^\beta \frac{\partial c_i^\beta}{\partial x} + f^\alpha c_i^\alpha v^\alpha + f^\beta c_i^\beta v^\beta\right),$$
$$i = 1, 2, 3. \qquad \text{(Eq 8)}$$

When the phase diagram, the tie lines and thermodynamic properties are known, the reactive diffusion path can be calculated.

The model allows for determination of the concentration of all of the components in both single-phase and two-phase regions. The schematic presentation is presented in Fig. 1.

In case of oxidation profile the α phase can be the alloy and β phase the oxide; thus, in α + β the two-phase region can be formed and the internal oxidation can be calculated. Note that in a single-phase region, for example α phase, the phase ratios are:

$$f^\alpha = 1 \quad \text{and} \quad f^\beta = 0;$$

thus, the mass conservation in α phase, Eq 8, can be rewritten in the following way:

$$\frac{\partial c_i^\alpha}{\partial t} = -\frac{\partial}{\partial x}\left(-D_i^\alpha \frac{\partial c_i^\alpha}{\partial x} + c_i^\alpha v^\alpha\right), \quad i = 1, 2, 3. \qquad \text{(Eq 9)}$$

Equation 9 is a standard mass conservation law for one phase. Moreover, when the sequence of the intermetallic phases is known, the thickness of the phases can be calculated.

The following steps should be implemented to solve the model:

The calculations performed at each time step:

(a) Solve volume continuity equation, Eq 7. Calculate diffusion flux, $D_i^j \frac{\partial c_i^j}{\partial x}$, and drift velocity, $v^j$, Eq 7;
(b) Solve the mass conservation law, Eq 8. Calculate concentrations, $\bar{c}_i$, $i = 1, 2, 3$, for advancement $t_k \rightarrow t_{k+1}$. Determine the conode which cross the $\bar{c}_i$, $i = 1, 2, 3$, in the two-phase region. The conodes are evenly distributed over the boundaries;
(c) Compute the intersection points of the conodes with the boundaries. These intersections determine the concentrations of the components in each phase, $c_i^\alpha$, $c_i^\beta$ for $i = 1, 2, 3$;
(d) Calculate the volume fractions of each phase, $f^\alpha$ and $f^\beta$;

The initial steps:

The remaining initial data necessary in the calculations include: terminal composition of the diffusion couple (representing the terminal points of the diffusion path), diffusion coefficients and molar volumes of the components in all phases, processing time and temperature.

## 4. Results and Discussion

Figure 2 shows the SEM/BSE image of the as-received microstructure of alloy IN 713C. The alloy exhibits a classical superalloy microstructure based on the γ-Ni matrix strengthened by γ'-Ni$_3$Al precipitates (Ref 1). Due to the high Nb content and relatively high carbon content, Nb carbides are observed (Fig. 2). The Nb-rich carbides were identified using SEM/EDX. The EDX spectra obtained from the carbide (Point 1) and the matrix (Point 2) shown in Fig. 2(b) show higher peak of niobium and slight peak of carbon. The Nb carbides formation is also one of the strengthening mechanisms (Ref 1). A similar microstructure of IN 713C has been reported by Binczyk and Śleziona (Ref 19) and Matysiak et al. (Ref 20). It was shown by Safari et al. (Ref 21) that the high-melting elements (e.g., Nb) segregate in dendritic region during the heat

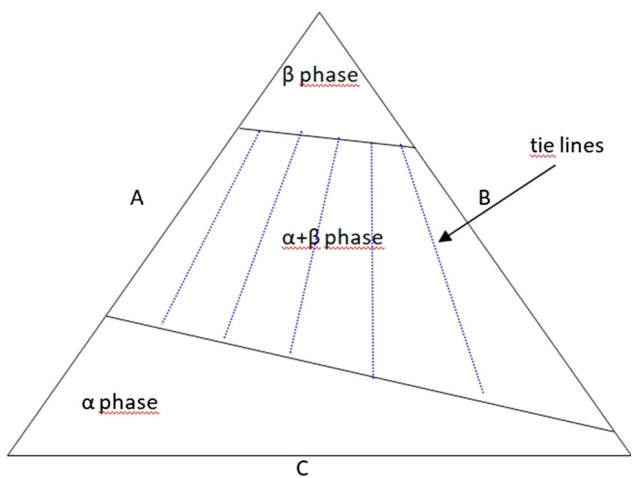

Fig. 1 Schematic presentation of the internal oxidation model

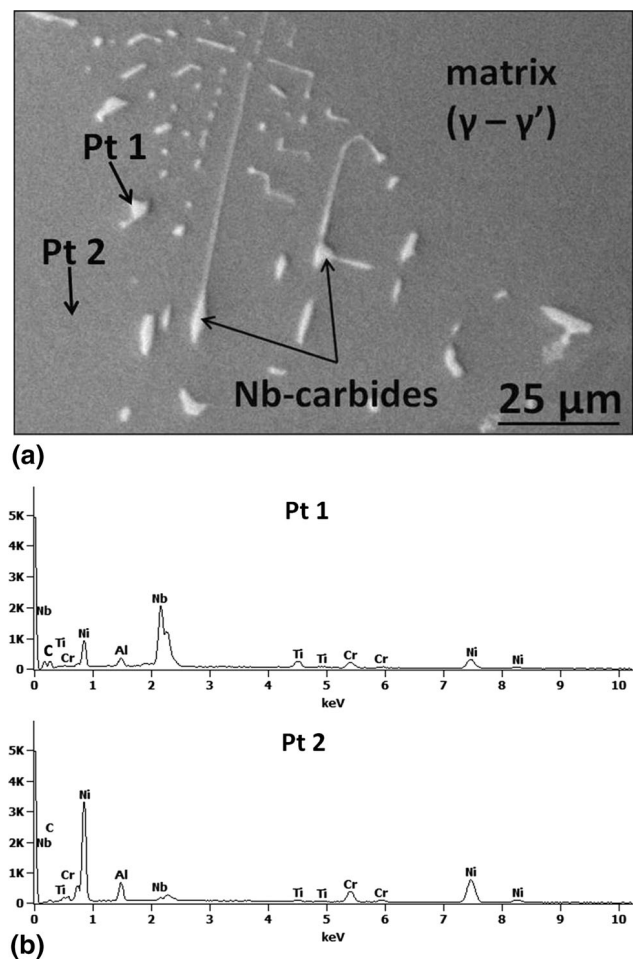

Fig. 2 (a) SEM/BSE image of the cross section of IN 713C in the as-received condition and (b) SEM/EDX spectra performed on carbide (Point 1) and matrix (Point 2) treatment. Matysiak et al. (Ref 20) found the presence of the primary carbides NbC in the interdenritic regions in the as-cast IN 713C. In the present work the investigated IN 713C was not heat-treated; therefore, Nb is present in the form of primary carbides in the eutectic region.

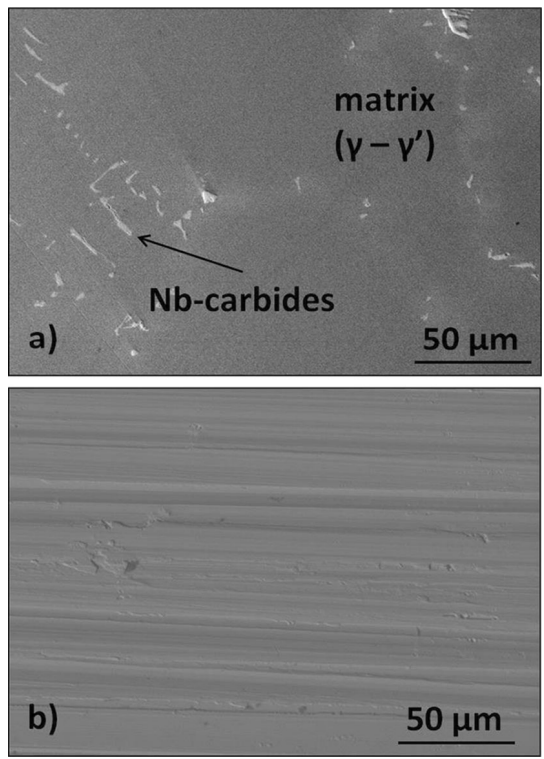

**Fig. 3** SEM/BSE images showing surface topography of IN 713C after (a) polishing up to 1-μm polishing and (b) grinding using 80-grit SiC paper

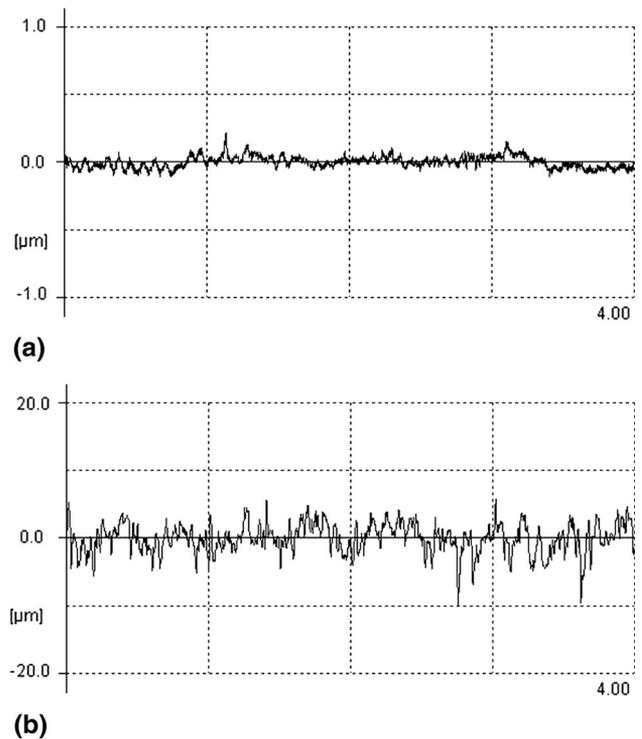

**Fig. 4** Roughness profiles of IN 713C after (a) polishing up to 1-μm polishing and (b) grinding using 80-grit SiC paper

**Table 1** Average values of surface roughness measured on polished and ground IN 713C

| Parameter | Polishing, 1 μm | | Grinding, 80 grit | |
|---|---|---|---|---|
| | Average value | SD | Average value | SD |
| $R_a$ | 0.024 | 0.001 | 1.186 | 0.22 |
| $R_z$ | 0.185 | 0.021 | 9.111 | 1.176 |
| $R_{max}$ | 0.227 | 0.038 | 12.469 | 3.044 |

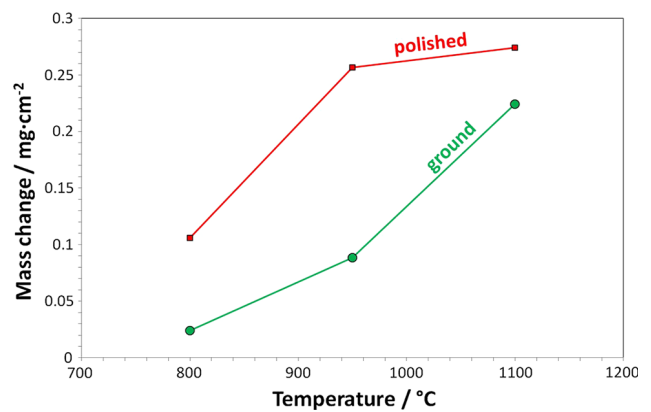

**Fig. 5** Mass changes plot obtained for polished and ground IN 713C exposed to air for 48 h at different temperatures

Figure 3(a) and (b) presents the surface topography registered using the SEM/BSE 3D mode. The images clearly reveal the difference between two surface preparation methods used in the present study. This observation is confirmed by the roughness profiles displayed in Fig. 4. The average $R_a$, $R_z$, $R_{z\ max}$ and $W_x$ values based on five measurements are summarized in Table 1. The measurements in Table 1 revealed one order of magnitude difference in roughness between the ground and polished specimens. It might be as well concluded that the ground specimen has a higher surface-to-volume ratio which provides a larger contact area between oxygen and the alloy surface, resulting in higher oxidation rates. Such an effect was previously observed by Eubanks et al. (Ref 22). It should be noted that the authors investigated the oxidation of pure iron. Moreover, the results in the present work indicate the opposite effect, i.e., grinding leads to a better oxidation behavior. The latter effect has a fundamentally different nature because in the case of the Ni-based superalloy grinding promotes scaling of a slow-growing alumina scale, resulting in lower net oxidation kinetics.

The mass change plot for the samples oxidized at 800-1100 °C is shown in Fig. 5. The mass gains of both ground and polished samples at 800 °C are about 0.02 and 0.11 mg cm$^{-2}$, respectively. At 950 °C the mass gain of the ground alloy specimen is 0.09 mg cm$^{-2}$, while the polished one exhibited the higher mass gain of 0.26 mg cm$^{-2}$. Finally, at 1100 °C the mass change of the ground and polished specimens is 0.22 and 0.27 mg cm$^{-2}$, respectively. From these results, one may conclude that the effect of surface preparation is most pronounced at lower temperatures. The effect gradually van-ishes with increasing temperature as the contribution of bulk diffusion to the outward transport of the scale-forming element exceeds the GB diffusion term.

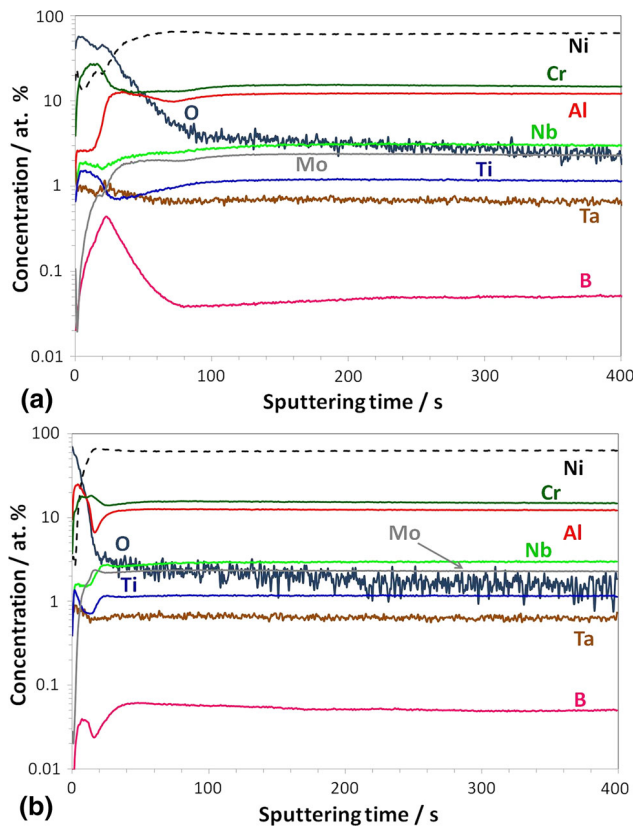

**Fig. 6** GD-OES depth profiles obtained for (a) polished and (b) ground IN 713C exposed in air for 48 h at 800 °C

### 4.1 Exposure to Air at 800 °C

The GD-OES depth profiles of polished IN 713C (Fig. 6a) revealed a thicker oxide scale compared to that formed on the ground specimen. In the outermost part of the oxide scale, a peak of Ni is observed, below which a co-enrichment of Cr and Ti is found. Below the Cr-rich zone a peak of Al is observed. The aluminum enrichment accompanied by lower oxygen concentration in the zone indicates the formation of internal $Al_2O_3$ precipitates. Between the outer scale and the internal oxidation zone, an enrichment of boron is detected. A similar enrichment at the bottom of the outer chromia scale was previously observed by Jalowicka et al. (Ref 23) and Nowak et al. (Ref 24), and it was associated with the formation of a $BCrO_3$ phase. The GD-OES depth profile for the ground IN 713C (Fig. 6b) showed a clear Al peak in the outer scale which is indicative of the presence of the $Al_2O_3$ scale. Contrary to the GD-OES depth profiles of the polished sample, no boron enrichment was detected below the outer Al-rich oxide scale.

The SEM images of the polished and ground (Fig. 7a and c, respectively) surfaces of IN 713C after exposure revealed different oxide scale morphologies formed on the dendritic and eutectic surface regions of the alloy. In the eutectic regions, Nb-rich oxides result from the oxidation of Nb carbides on both polished and ground surfaces. However, it should be noted that the amount of Nb-rich oxides in the eutectic regions is higher on the polished samples. Apparently, grinding affects the oxidation mechanism leading to the formation of protective $Al_2O_3$, which in turn suppresses the formation of Nb-rich oxides. The SEM images of the cross-sectioned samples revealed that the polished sample (Fig. 7b) formed a thicker oxide scale consisting of a Ni-/Cr-rich oxide on top of the oxide scale accompanied by an Al-rich sublayer. As it can be seen in Fig. 7(d), Ni/Cr mixed oxides are locally present. The SEM results are in very good agreement with the GD-OES depth profiles.

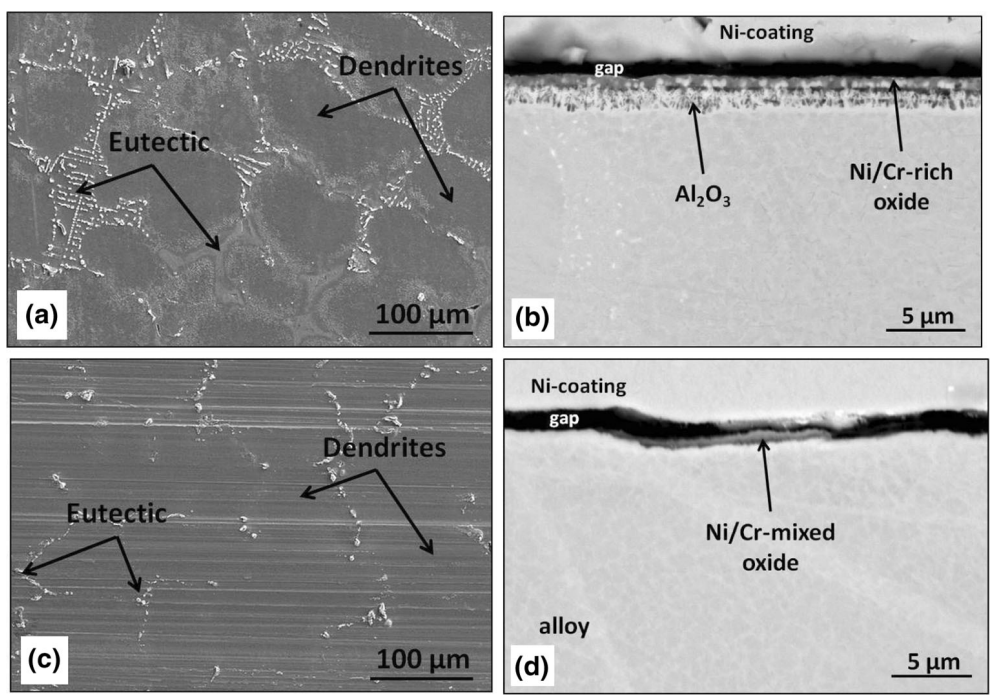

**Fig. 7** SEM/SE images showing surface and cross sections of polished (a) and (b) and ground (c) and (d) IN 792 after isothermal oxidation test at 800 °C for 48 h in air

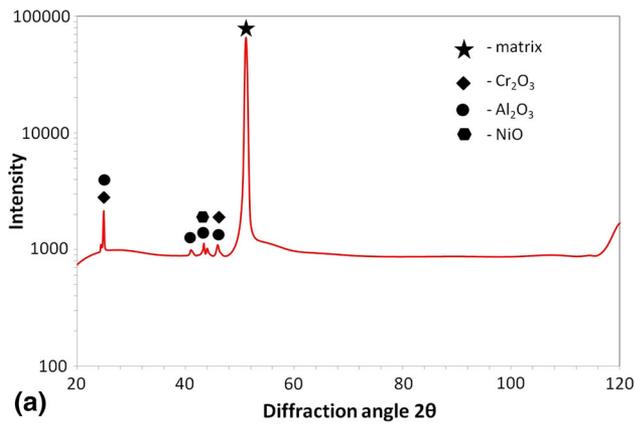
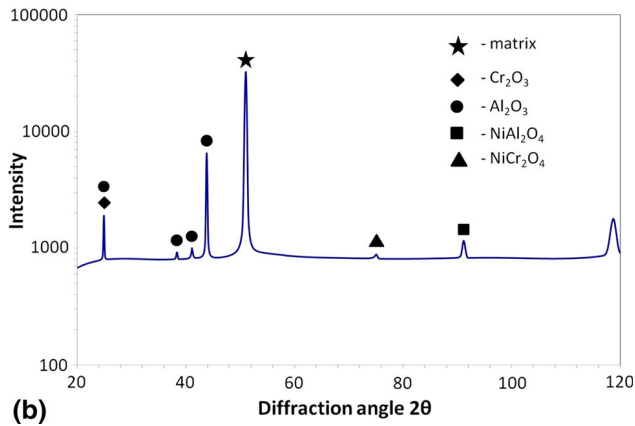

**Fig. 8** XRD patterns obtained from (a) polished and (b) ground IN 713C oxidized at 800 °C for 48 h in air

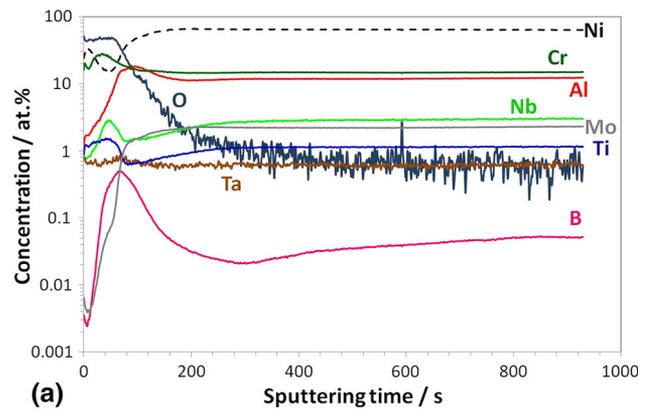
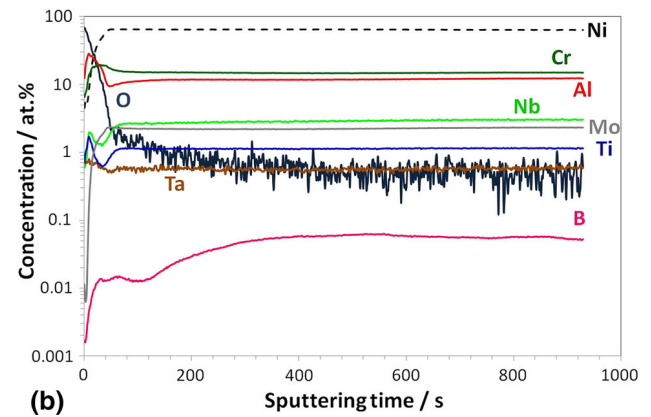

**Fig. 9** GD-OES depth profiles obtained for (a) polished and (b) ground IN 713C exposed in air for 48 h at 950 °C

The phase analysis performed using XRD diffractometer showed the presence of $Cr_2O_3$, $Al_2O_3$ and NiO on the surface of the polished sample (Fig. 8a). The XRD patterns obtained from ground sample after oxidation revealed the presence of $Cr_2O_3$, $Al_2O_3$; however, in contrast to the observation on polished sample, no peak from NiO is present. Instead of NiO, the presence of $NiAl_2O_4$ and $NiCr_2O_4$ spinels is detected (Fig. 8b). The most likely spinels originated from the transient stage of oxidation (Ref 25). Moreover, higher intensity of $Al_2O_3$ peak obtained on ground as compared to polished sample after exposure is observed. The latter result in combination with GD-OES depth profiles obtained on ground sample (Fig. 6b) indicates the formation of very thin, continuous alumina layer.

### 4.2 Exposure to Air at 950 °C

The GD-OES depth profiles of the polished IN 713C sample (Fig. 9a) showed that the outer part of the oxide scale is rich in Ni. Below the outer Ni peak, a co-enrichment in Cr and Nb is found. Furthermore, the entire oxide scale is slightly enriched in Ti. Similar to observation at 800 °C (Fig. 6), an enrichment of boron is detected at the oxide scale–alloy interface in the polished specimen. Below the boron peak, a zone of boron depletion is present. Additionally, an internal zone oxidation zone of Al was found also at 950 °C.

The GD-OES depth profiles of the ground specimen revealed an aluminum peak in the outer part of the oxide scale accompanied by a slight enrichment in Ti and Nb (Fig. 9b). Similar to observation at 800 °C, no boron peak is observed.

An SEM surface analysis of the polished and ground samples (Fig. 10a and d, respectively) revealed a relatively uniform oxide scale formed on the polished sample, while the formation of mixed oxides in the dendrite and eutectic regions of the alloy microstructure is observed for the ground samples. Considering the cross sections (Fig. 10c and f), one can see an outer Cr-rich oxide scale and internally oxidized aluminum. On the contrary, the polished sample developed a continuous and thin aluminum-rich oxide scale. However, the alumina-based oxide scale in Fig. 10(f) is not microstructurally uniform containing the dense and flat oxide scale regions as well as oxides spikes. The former is typical for the $\alpha$-$Al_2O_3$, while the spikes are characteristic of $\theta$-$Al_2O_3$. It is known from the literature that a metastable $\theta$-$Al_2O_3$ phase can form at 950 °C (Ref 26). The nodules observed on the surface of oxidized ground IN 713C (Fig. 10d) might be correlated with $\theta$-$Al_2O_3$. The identification of the $\theta$-$Al_2O_3$ formation might be supported by the larger zone of $\gamma'$-depletion below the spike-shaped oxides which in turn is associated with the higher growth rate of $\theta$-$Al_2O_3$ compared to $\alpha$-$Al_2O_3$ due to more rapid cation diffusion in these structures as compared to $\alpha$-$Al_2O_3$ (Ref 27). The Al consumption below the $\theta$-$Al_2O_3$ scale is faster, resulting in a deeper $\gamma'$-depleted zone. It was established that the transient $\theta$-$Al_2O_3$ eventually transformed into $\alpha$-$Al_2O_3$ after some exposure time. At higher temperatures, the transformation time was shorter. The authors found a typical "whisker topography" of $\theta$-$Al_2O_3$ which is qualitatively similar to the oxide morphology observed at 950 °C on the ground specimen in the present work (Fig. 11).

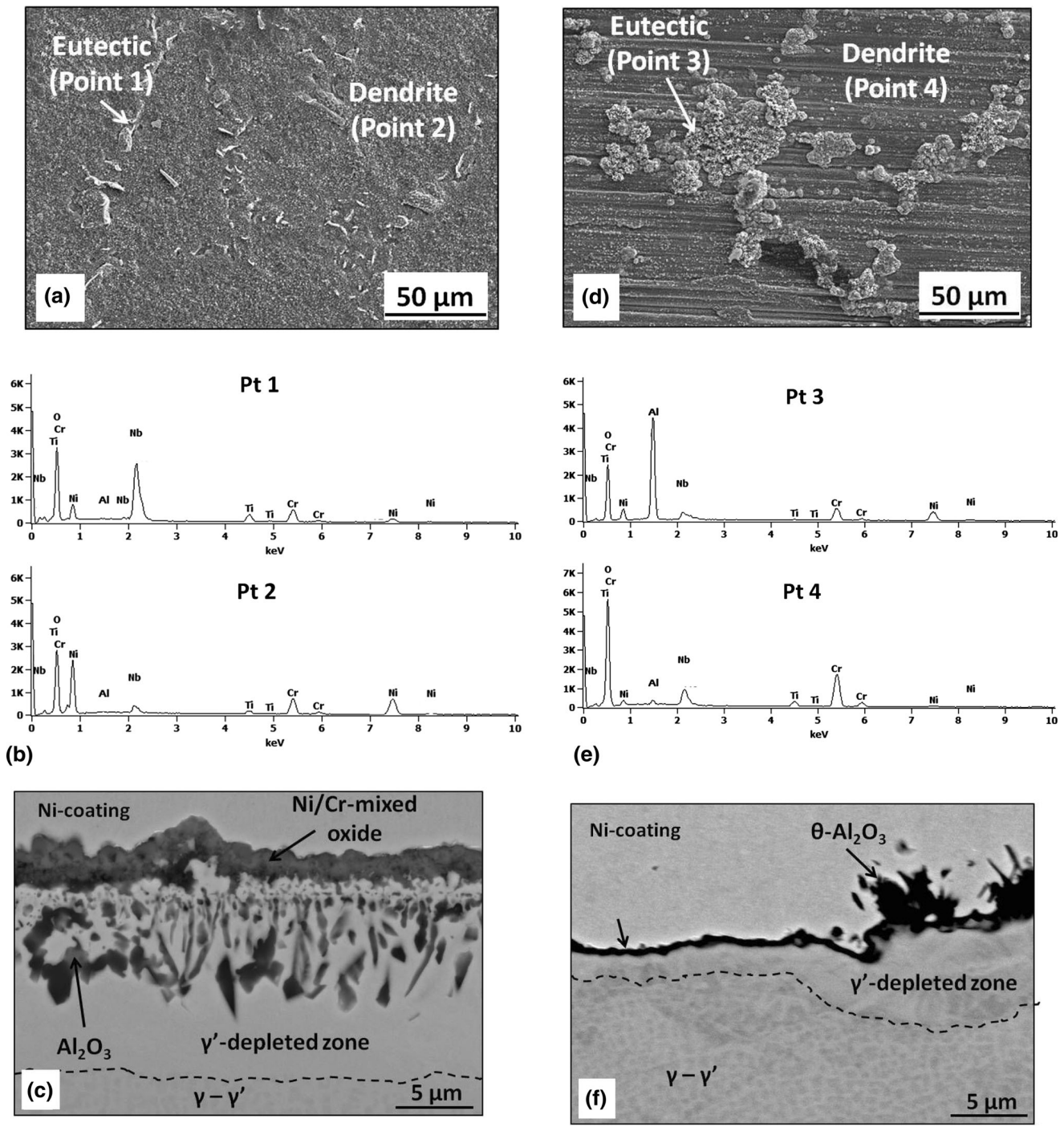

**Fig. 10** SEM images showing surfaces, SEM-EDX spectra and cross sections of the polished (a-c) and ground (d-f) IN 792 after isothermal oxidation test at 950 °C for 48 h in air. The dashed line indicates the zone of γ′ depletion

Figure 12 shows the XRD patterns obtained from polished and ground IN 713C after air oxidation at 950 °C. On polished surface, formation of $Cr_2O_3$, $Al_2O_3$, NiO and small reflexes from $NiCr_2O_4$ is observed (Fig. 12a), while on ground surface formation of $Cr_2O_3$, $Al_2O_3$ and $NiAl_2O_4$ and $NiCr_2O_4$ spinels instead of NiO was found (Fig. 12b). Similar to the observation at 800 °C, the intensity of $Al_2O_3$ peaks is higher for ground surface in comparison with polished one. The latter indicates the formation of relatively pure alumina scale.

### 4.3 Exposure of IN 713C to Air at 1100 °C

The GD-OES depth profiles of the polished and ground IN 713C specimens after air exposure at 1100 °C (Fig. 13a and b) revealed similar oxidation behavior as at the lower temperatures. The oxide scale on the polished sample is enriched in Ni and Cr in the outer part. In contrast to the polished sample oxidized at lower temperatures, the peak of Al is not correlated with the lower concentration of oxygen which is an indication

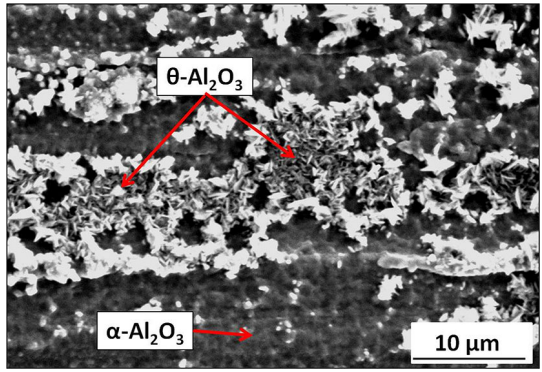

**Fig. 11** SEM/SE image showing typical whisker structure of θ-$Al_2O_3$ formed on the surface of IN 713C exposed at 950 °C for 48 h in air

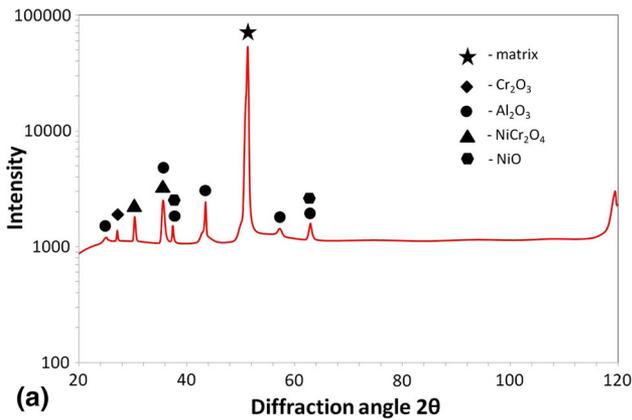

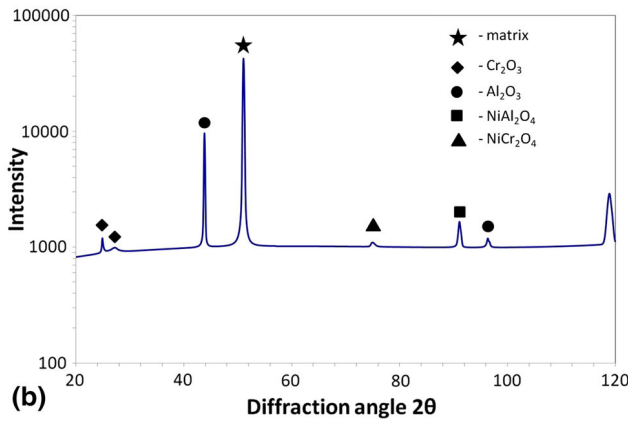

**Fig. 12** XRD patterns obtained from (a) polished and (b) ground IN 713C oxidized at 950 °C for 48 h in air

of the continuous $Al_2O_3$ sublayer underneath the Ni-/Cr-rich oxide (Ref 17).

The GD-OES depth profiles for the ground sample revealed the formation of an external Al-rich oxide (Al enrichment in the outer part of the oxide scale). However, a slight enrichment in Ni, Cr, Nb and Ti is also observed at the outermost part of the oxide scale, suggesting the formation of a spinel layer during the transient stage of exposure (Ref 5, 28).

The results of the SEM surfaces analyses of oxidized samples (Fig. 14a and c) in combination with the SEM cross section images (Fig. 14b and d) confirmed the GD-OES results, i.e., the outer part of the oxide scale in polished samples consists of Ni-/Cr-rich oxide and a sublayer of $Al_2O_3$, while the ground sample developed a continuous $Al_2O_3$ oxide scale with local intrusions of Ni/Cr mixed oxide.

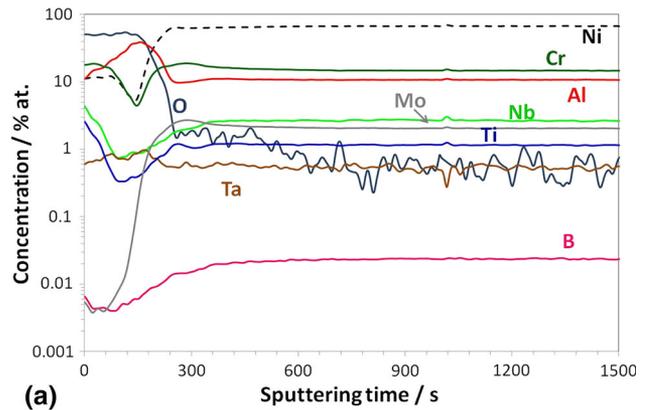

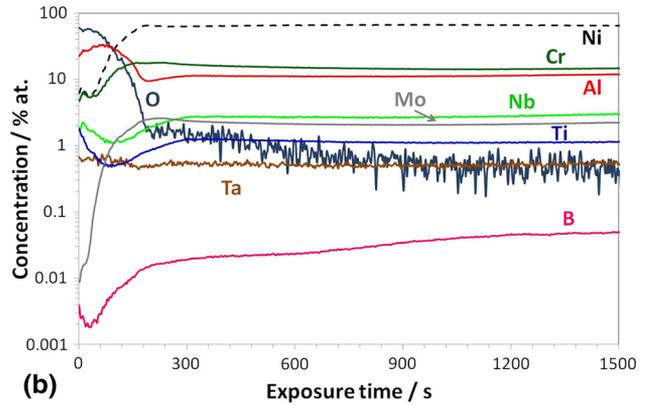

**Fig. 13** GD-OES depth profiles obtained for (a) polished and (b) ground IN 713C exposed in air for 48 h at 1100 °C

### 4.4 Summary of the Results

The polished alloy samples exposed to air at 800 and 950 °C were unable to develop a fully continuous alumina layer. The GD-OES depth profiles revealed a boron enrichment at the scale–alloy interface. However, a thin alumina scale is formed on the ground samples oxidized at both temperatures. Furthermore, the GD-OES depth profiles for the ground IN 713 specimens obtained at 800 and 950 °C demonstrated no boron enrichment below the outer oxide scale. A similar effect of boron diffusion inhibition through the alumina scale was previously reported by Nowak et al. (Ref 24). For both polished and ground samples exposed to air at 1100 °C, no boron enrichment was found. However, it is noteworthy that a continuous sublayer of alumina was observed for both samples and therefore boron enrichment is not expected.

According to the oxidation map by Giggins and Pettit (Ref 3), IN 713C is classified as a marginal alumina former, if only the ternary Ni-Cr-Al alloy is considered. However, the presented results indicate that the oxide scale morphology can be also affected by surface treatment. When ground, the alloy specimens formed a protective, continuous alumina layer at all tested temperatures, while the polished specimens developed an outer Ni/Cr mixed oxide accompanied by an internal oxidation zone of aluminum. Therefore, it can be concluded that an increase in surface roughness by one order of magnitude

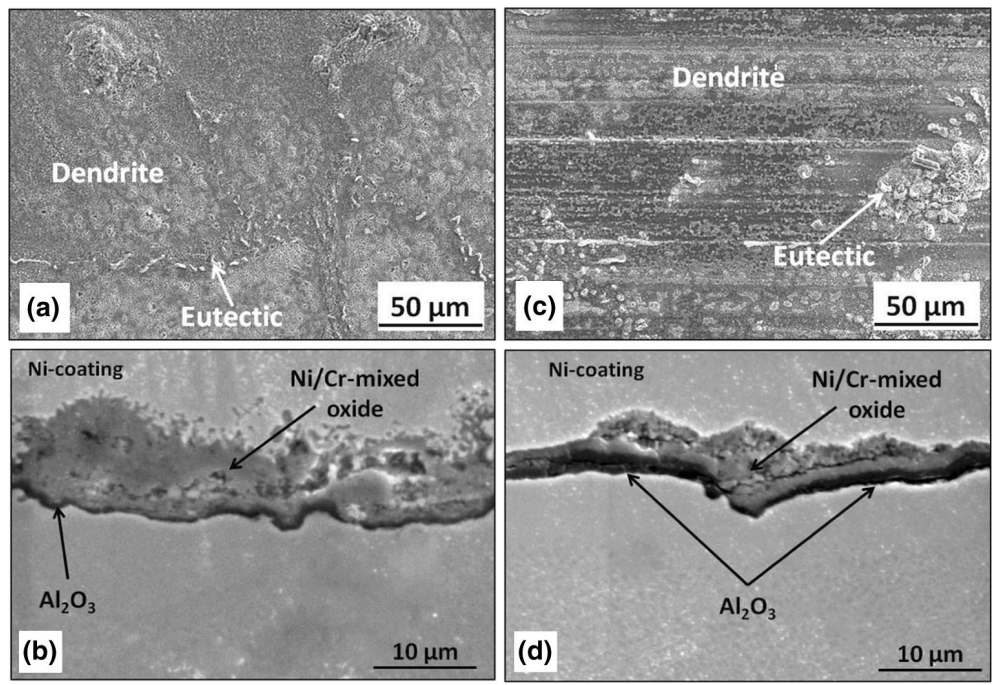

**Fig. 14** SEM/SE images showing surface and cross sections of polished (a) and (b) and ground (c) and (d) IN 792 after isothermal oxidation test at 1100 °C for 48 h in air

promoted the formation of a protective alumina scale. It is known that at lower temperatures grain boundary diffusion dominates over lattice diffusion as a result of the lower activation energy for the grain boundary processes due to more disordered structures in the boundaries (Ref 28). Considering the latter, it is suggested that surface preparation affects the amount and type of defects which might additionally lower the activation energy for diffusion in the cold-worked region. It should also be noticed that surface treatment drastically alters the oxide scale morphology, while the alloy is generally assumed to be an alumina former.

## 5. Proposed Oxidation Mechanisms

Based on the obtained results, the following mechanism of oxide scale formation for the ground and polished specimens at lower temperatures is proposed. The material in the as-received condition is qualitatively represented in Fig. 15 as "as-received condition." The blue, green and red dots represent Ni, Cr and Al atoms, respectively. These atoms are randomly distributed over the entire volume of the specimen. From the material in the as-received condition, samples were prepared in two different ways: polishing up to 1-μm silica grains and grinding with 80-grit SiC paper. It is proposed that the different surface preparation results in the introduction of different types and number of defects into the near-surface region (Fig. 15 stage 1). The defects are marked with black dots and lines in Fig. 15. During exposure to air at a high temperature, the atoms start to diffuse as shown in Fig. 15 in stage 2. The near-surface region of the ground specimen is characterized by a higher concentration of defects. These defects accelerate the diffusion of the scale-forming element, which in the present case is Al. On the contrary, the polished surface contains only minor amounts of defects; therefore, the number of easy paths for Al diffusion is limited. Thus, in the case of the polished sample aluminum needs to diffuse within the grain volume. The latter results in the formation of Ni/Cr mixed oxide, while the formation of a relatively pure alumina scale on ground surface is determined (as shown in stage 3). After a longer exposure time, a thin alumina sublayer is formed as shown at stage 4 in Fig. 15. As mentioned in the previous section, such an effect is being suppressed with increasing temperature.

## 6. Summary and Conclusions

The effect of surface preparation on the oxide scale formation has been studied in the present study and the following conclusions are drawn:

- Polished samples develop an outer Ni-/Cr-rich oxide along with a zone of internal aluminum oxidation, while the ground samples produce a protective external alumina;
- IN 713C exposed to air at 950 °C tends to form $\theta$-$Al_2O_3$;
- Decreasing the surface roughness by two orders of magnitude results in the formation of Ni/Cr mixed oxide scale on a nominally alumina-forming alloy, the effect being more pronounced at lower temperatures;
- The oxide scale morphology was shown to be strongly affected by the surface preparation methods such as grinding and/or polishing. The mechanical deformation of the near-surface region results in a higher local density of defects and faster diffusion of Al toward the alloy surface thus promoting protective scaling.

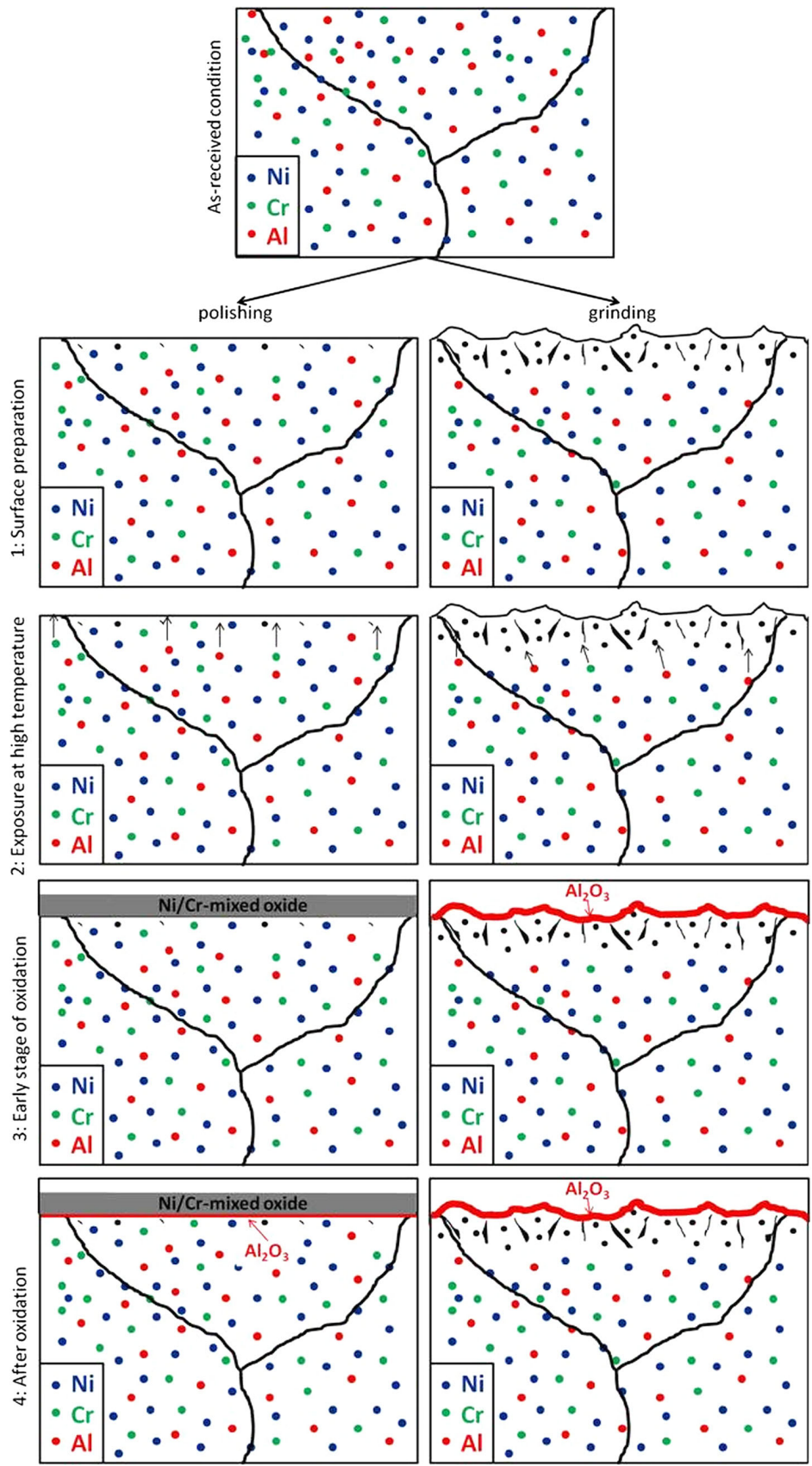

Fig. 15 Proposed mechanisms of oxide scale formation for the alloy with polished and ground surfaces at 800 °C

## Acknowledgment

This project is financed within the Marie Curie COFUND scheme and POLONEZ program from the National Science Centre, Poland, POLONEZ Grant No. 2015/19/P/ST8/03995. This project has received funding from the European Union's Horizon 2020 Research and Innovation Programme under the Marie Skłodowska Curie Grant Agreement No. 665778.

## Open Access